\pdfoutput=1

\documentclass[%
reprint,
 amsmath,amssymb,
 aps,
 prd,
 10pt,
]{revtex4-2}

\DeclareUnicodeCharacter{2212}{-}
\DeclareUnicodeCharacter{03BC}{\mu}
\usepackage{graphicx}
\usepackage{dcolumn}
\usepackage{bm}
\usepackage{hyperref}
\usepackage[utf8]{inputenc}

\usepackage{subfig}
\usepackage{ragged2e}
\usepackage[compat=1.1.0]{tikz-feynman}
\usepackage{color}
\definecolor{comment}{rgb}{.4,.4,.4}
\usepackage{algorithm}
\usepackage{subcaption}
\usepackage[
noEnd=false,
indLines=false,
italicComments=false,
rightComments=false,
commentColor=comment,
beginComment={~~//~},
endComment={},
]{algpseudocodex}
\algrenewcommand\textproc{\textrm}
\newcommand\hrulex[1]{\rule{\columnwidth}{#1}}
\newcommand\urulex[1]{\vspace{+.5\baselineskip}\hrulex{#1}\vspace{-.5\baselineskip}}
\newcommand\drulex[1]{\vspace{-.5\baselineskip}\hrulex{#1}}

\begin{document}

\preprint{APS/123-QED}

\title{Search for light Dark Sectors with GeV Muon Beams}%

\author{Zijian \surname{Wang}}%
\email{wangzijian@stu.pku.edu.cn}%
\author{Leyun \surname{Gao}}%
\author{Zhuo \surname{Chen}}%
\author{Cheng-en \surname{Liu}}%
\author{Jinning \surname{Li}}%
\author{Qite \surname{Li}}%
\author{Chen \surname{Zhou}}%
\author{Qiang \surname{Li}}%
\affiliation{School of Physics and State Key Laboratory of Nuclear Physics and Technology, Peking University, Beijing, 100871, China}%

\author{Yu \surname{Xu}}

\author{Xueheng \surname{Zhang}}

\author{Liangwen \surname{Chen}}

\author{Zhiyu \surname{Sun}}
\affiliation{Institute of Modern Physics, Chinese Academy of Science, Lanzhou, 730000, China;\\University of Chinese Academy of Sciences, Beijing, 100049, China}

\author{Ce \surname{Zhang}}
\email[]{ce.zhang@liverpool.ac.uk}
\affiliation{University of Liverpool, Liverpool, United Kingdom}

\date{\today}

\begin{abstract}

Sub-GeV light dark matter often requires new light mediators, such as a dark $Z$ boson in the $L_\mu - L_\tau$ gauge theory. We study the search potential for such a $Z^\prime$ boson via the process $\mu e^- \to \mu e^- X$, with $X$ decaying invisibly, in a muon on-target experiment using a high-intensity 1–10 GeV muon beam from facilities such as HIAF-HIRIBL. Events are identified by the scattered muon and electron from the target using the silicon strip detectors in a single-station telescope system. Backgrounds are suppressed through a trained BDT classifier, and activity in downstream subdetectors remains low. We find that this approach can probe a $Z^\prime$ in the 10 MeV mass range with improved sensitivity. Nearly three orders of magnitude improvement is achievable using a full multi-telescope station system using a 160 GeV muon beam at CERN, such as in the MUonE experiment.

\end{abstract}

\maketitle


\section{Introduction}

Minimal scenarios with light (sub-GeV) dark matter (DM) whose relic density is obtained from thermal freeze-out must include new additional dark sectors (DS) including some light mediators. A very well-motivated example is a new “dark” massive vector gauge boson mediator~\cite{Battaglieri:2017aum}. Many such scenarios~\cite{PhysRevD.104.115008, Agrawal_2014, Kamada_2018, Bickendorf_2022} also suggest new feeble interactions with muons mediated by scalar, pseudoscalar, or vectorlike particles. 

A simple yet successful class of such models is based on the $L_\mu - L_\tau$ gauge theory~\cite{Foot:1990mn, He:1990pn, He:1991qd, Foot:1994vd}, where a massive dark gauge boson exists associated with this gauge symmetry, called the $L_\mu - L_\tau$ gauge boson ($Z^\prime$), where $L_\mu$ and $L_\tau$ are the $\mu$ and $\tau$ lepton numbers, respectively. 

In the $L_\mu - L_\tau$ {\bf vanilla} model, the $Z^\prime$ vector boson from the broken $U(1)_{L_\mu-L_\tau}$ symmetry couples directly to the second and third lepton generations, and their corresponding left-handed neutrinos~\cite{He:1990pn,He:1991qd,Foot:1994vd,Altmannshofer:2016jzy,Kile:2014jea,Park:2015gdo}. The corresponding Lagrangian is as follows~\cite{NA64:2024nwj}:
\begin{equation}
\label{eq:lagrangian}
\mathcal{L}\supset-\frac{1}{4}F_{\alpha\beta}^\prime F^{\alpha\beta\prime}+\frac{m_{Z^\prime}^2}{2}Z_\alpha^\prime Z^{\alpha\prime}-g_{Z^\prime}Z_\alpha^{\prime}J_{\mu-\tau}^\alpha,
\end{equation}
where $J_{\mu-\tau}^\alpha$ is the $U(1)_{L_\mu-L_\tau}$ leptonic current, 
\begin{equation}
    J_{\mu-\tau}^\alpha=\big(\bar{\mu}\gamma^\alpha\mu-\bar{\tau}\gamma^\alpha\tau
        +\bar{\nu}_\mu\gamma^\alpha P_L\nu_\mu
        -\bar{\nu}_\tau\gamma^\alpha P_L\nu_\tau\big),
\end{equation}
and $F_{\alpha\beta}^\prime$ the field strength tensor associated with the massive vector field $Z_\alpha^{\prime}$, $g_{Z^\prime}=\epsilon_{Z^\prime}e$ the coupling of $Z^{\prime}$ to SM particles, and $P_L$ the left-handed chiral projection operator. 

Within this {\bf vanilla} model, the gauge boson $Z^\prime$ decays invisibly to SM neutrinos, such that 
\begin{equation}
    \label{eq:decays-neutrinos}
    \Gamma(Z^\prime \rightarrow\overline{\nu}{\nu})=\frac{\alpha_{Z^\prime}m_{Z^\prime}}{3},
\end{equation}
with $\alpha_{Z^\prime}=g_{Z^\prime}^2/(4\pi)$. For $m_{Z^\prime}>2m_\mu$, the visible decays to SM leptons, $Z^\prime\rightarrow\bar{\mu}\mu$, become kinematically allowed.

The {\bf vanilla} model can also be extended to include DM ($\chi$),  interacting with $Z^\prime$ through $\mathcal{L}\supseteq -g_\chi Z_\alpha^\prime J_\chi^\alpha$, with $J_\chi^\alpha$ a DS current as~\cite{Kahn:2018cqs}:
\begin{equation}
    J_{\chi}^{\alpha}=g_{\chi}
    \begin{cases}
        i\chi^{\ast}\partial^\alpha\chi+\text{h.c.}, & \text{complex scalar} \\
        1/2\overline{\chi}\gamma^\alpha\gamma^{5}\chi, & \text{Majorana}\\
        i\overline{\chi}_1\gamma^\alpha\chi_2, & \text{pseudo-Dirac} \\
        \overline{\chi}\gamma^\alpha\chi, & \text{Dirac}\\
    \end{cases},
\end{equation}
with $g_\chi$ as the coupling of $Z^\prime$ to the DM candidates. More details including those on the relic density can be found in ref.~\cite{Kahn:2018cqs,NA64:2024nwj}. For example,  in the $m_{Z^\prime}  \gg m_\chi$ for a Dirac $\chi$ particle. Defining
the dimensionless variable $y \equiv g_\chi^2 g_{\mu-\tau}^2 (m_\chi/m_{Z^\prime})^4$, the cross section and abundance 
have the approximate scaling  
\begin{equation}
\begin{split}
& \langle  \sigma v \rangle \simeq \frac{3 g_\chi^2 g^2 m_\chi^2}{\pi^2 m_{Z^\prime}^2} = \frac{3y}{\pi m_\chi^2},   
\\
~ \implies ~
&\Omega_\chi h^2 \sim 0.1 \left( \frac{3 \times 10^{-9}}{ y}\right)   \biggl( \frac{m_\chi}{ \rm GeV}\biggr)^2.
\end{split}
\end{equation}

This dark boson, $Z^\prime$, has been sought at the LHC~\cite{CMS:2018yxg}. In particular, analyses of $Z \to 4\mu$ events at $\sqrt{s}=13$ TeV have placed stringent bounds on $Z^\prime$ bosons with masses of order 1–100 GeV, showing no significant deviation from the Standard Model expectation and thus constraining the allowed $Z^\prime$ coupling strength. It could also be produced in the bremsstrahlung-like reaction of GeV scale muons with a target (nuclei or electron) followed by its subsequent invisible decay, $\mu N \to \mu N Z^\prime$ or $\mu e^- \to \mu e^- Z^\prime$, where $Z^\prime$ decays invisibly into neutrinos or dark matter. Examples include recent experimental NA64 results~\cite{NA64:2024klw,NA64:2024nwj} and a proposed idea based on MUonE platform~\cite{Asai:2021wzx}, both rely on the M2 beamline at the CERN Super Proton Synchrotron with a momentum of 160~GeV/$c$ ($\mathcal{O}(100~\mathrm{GeV})$). 

Here we extend the study based on the MUonE experiment~\cite{Asai:2021wzx}, which employs a multi-station telescope system within the {\bf vanilla} model, while focusing on 1–10~GeV muon beams from the HIAF Fragmentation Separator (HIRIBL) at the High Intensity Heavy-ion Accelerator Facility (HIAF). The lower center-of-mass (COM) energy brings enhanced sensitivity to $Z^\prime$ in the low-mass region at the level of 10~MeV. By combining HIRIBL's high-intensity, tunable muon beam with the PKMu setup, our approach enables precise reconstruction of scattering events and effective background suppression. This configuration enhances sensitivity to light $Z^\prime$ bosons in the low-mass region, enabling precise exploration of dark sector physics.

\section{The joint HIRIBL-PKMu Experiment}\label{sec:pkmuon}

HIAF~\cite{xu_feasibility_2025, an2025highprecisionphysicsexperimentshuizhou} is a cutting-edge scientific facility under construction in Huizhou, China, designed to provide high-energy, high-intensity ion beams for a wide range of research applications, including nuclear physics, atomic physics, and muon beam generation. The facility is capable of delivering ion beams ranging from protons to uranium, with energies on the order of GeV/u and intensities as large as $10^{11} \sim 10^{12}$ particles per pulse (ppp). HIAF includes a superconducting ion linac (iLinac), a synchrotron booster ring (BRing), and a high-energy fragment separator (HIRIBL), which can filter out unwanted particles, such as pions, neutrinos, and electrons, provide a high-purity muon beam for use in applications such as muon tomography and fundamental particle physics experiments.

\begin{figure}[H]
	\centering
	\includegraphics
	[width=0.45\textwidth]{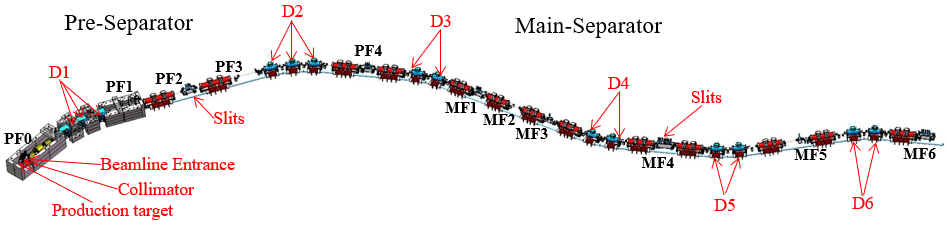}
	\caption{Schematic layout of the HIRIBL.}
	\label{fig:HFRS}
\end{figure}

As FIG.\ref{fig:HFRS} shows, the HIRIBL consists of two main stages: the Pre-Separator and the Main-Separator. The Pre-Separator, which extends from PF0 to PF4, is responsible for eliminating the primary beam and other undesired fragments created during the interaction process. This stage is crucial for reducing contamination in the beam before it moves on to the Main-Separator. The Main-Separator, which spans from PF4 to MF6, plays a more advanced role in purifying the beam by using powerful magnetic and electrostatic fields to separate particles based on their mass-to-charge ratio. This separation is essential for isolating muons from the mix of secondary particles generated during the target interaction. 

The HIRIBL is designed to deliver a highly purified muon beam by precisely manipulating high-momentum secondary particles. Operating with a magnetic rigidity of up to 25 Tm, it can effectively select particles with momenta up to 7.5 GeV/$c$. The strong magnetic field and advanced beamline optics enable accurate momentum selection and particle separation. Beam purification is further enhanced through dedicated particle identification elements, such as magnetic spectrometers and absorbers, which exploit the distinct mass and charge differences between muons and other secondary particles. Together, these features ensure that only muons are retained in the final beam, making it well suited for high-precision experimental applications.

The ability to produce muons with both high intensity and high purity is a key feature of the HIRIBL. For example, the maximum muon yield for the $\mu^+$ beam can reach $8.2 \times 10^6 \mu/s$ with a purity of approximately 2\% at a momentum of 3.5 GeV/c, while the $\mu^-$ beam achieves a yield of $4.2 \times 10^6 \mu/s$ with a purity of approximately 20\% at 1.5 GeV/c. These results show that the HIRIBL is capable of generating high-intensity muon beams. When appropriate purification strategies are used, a purity close to 100\% can be achieved, making the muon beam highly suitable for applications that require high precision, such as muon tomography, which benefits from a clean signal free from background noise.



Ref.~\cite{Yu:2024spj} proposed a new experimental scheme, named PKMu, to directly detect light mass dark matter through its scattering with abundant atmospheric or accelerator muons. The initial plan involves using free cosmic-ray muons interacting with dark matter in a volume surrounded by tracking detectors to trace potential interactions between muons and dark matter. Such a proposal can be extended to cover many other physics cases, such as those involving dark bosons. 

\begin{figure}[H]
\includegraphics[width=1.0\columnwidth]{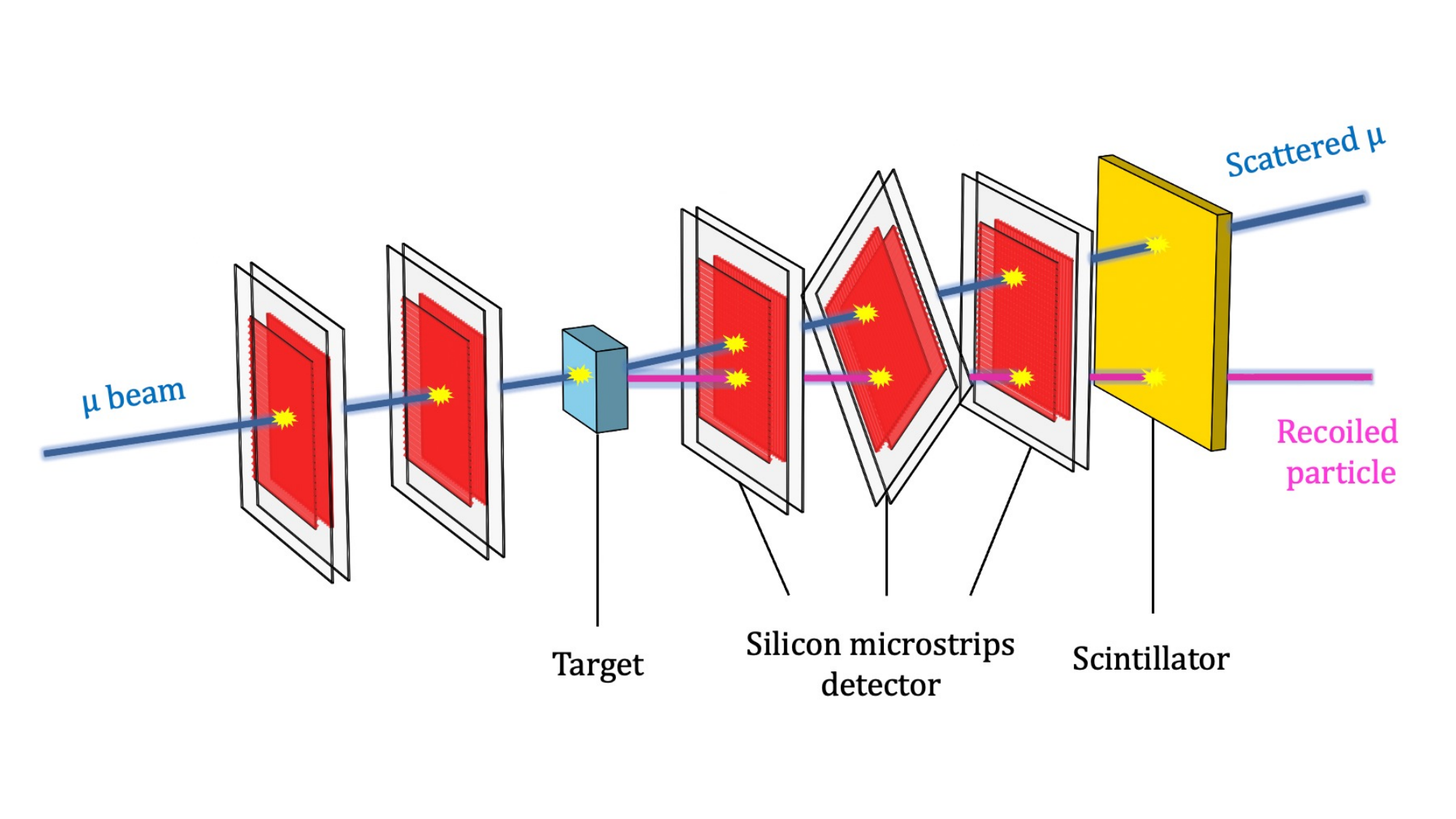}
\caption{\justifying A proposed muon experiment to probe $Z^\prime$ through muon on-target collisions.}
\label{fig:pkmuonz}
\end{figure}

Accordingly, we implement a PKMu-inspired setup using the HIAF-HIRIBL muon beam. FIG.~\ref{fig:pkmuonz} shows the schematic for a proposed muon experiment to probe $Z^\prime$ through muon on-target collisions. 
The schematic setup consists of several functional components: tracking detectors (shown in red), such as silicon strip detectors, which are well suited for detecting charged particles (muons and electrons); a target (shown in blue) containing electrons for collisions with cosmic muons or muon beams; and a particle identification (PID) system (shown in yellow), which may include scintillators and, optionally, an electromagnetic calorimeter (ECAL) for measuring the energy of electrons.
This system can help distinguish muons from electrons and further suppress Standard Model background contributions.

\section{Event Generation and Simulation}\label{sec:signal}

MadGraph5\_aMC@NLO 3.5.5~\cite{Alwall:2014hca} is used to simulate $\mu e^- \to \mu e^- Z^\prime$ for varying muon energies $E_\mu$ and $Z^\prime$ masses $m_{Z^\prime}$. The $Z^\prime$ decay width $\Gamma_{Z^\prime}$ scales with $m_{Z^\prime}$, and events generated with $g_{Z^\prime}=1$ can be reweighted to arbitrary couplings via a global cross-section factor.

For a given $E_\mu$, the two nearest grid points ($E_{\mu1}$, $E_{\mu2}$) are identified, and the corresponding cross sections ($\sigma_1$, $\sigma_2$) and histogram templates ($H_1$, $H_2$) are retrieved. A linear interpolation with energy-dependent weights $w_1$ and $w_2$ is performed in both cross section and histogram space, and a random draw according to $w_1$ selects the histogram to sample from, yielding the outgoing momenta $\vec{p}_\mu$, $\vec{p}_e$, and $\vec{p}_{\text{miss}}$ in the lab frame. This procedure, summarized in Algorithm~\ref{alg:interp}, then transforms the event kinematics into the COM frame, where the scattering angles $\alpha$ and $\phi$ are sampled from the interpolated distribution. The outgoing muon and electron three-momenta are boosted back to the laboratory frame using the Lorentz parameters $\gamma$ and $\beta$ determined by $E_\mu$, and the event is rotated to align the $z$-axis with the incoming muon direction. This yields the final momenta $\vec{p}_{\mu\,\text{out}}$ and $\vec{p}_{e\,\text{out}}$, ready for storage or further analysis. The procedure preserves angular distributions and cross-section continuity while significantly accelerating large-scale event generation.

\begin{figure}[t]
\centering
\subfloat[lab frame]{\label{fig:lab-frame}\includegraphics[width=.5\columnwidth]{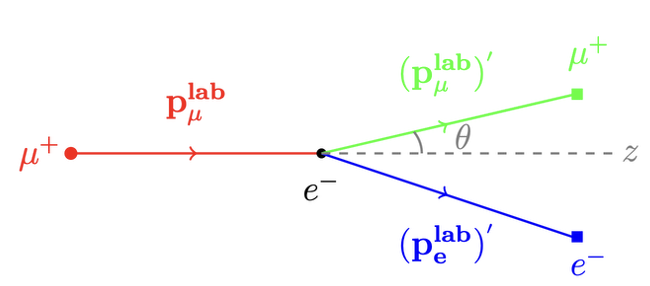}}
\subfloat[COM frame]{\label{fig:com-frame}\includegraphics[width=.5\columnwidth]{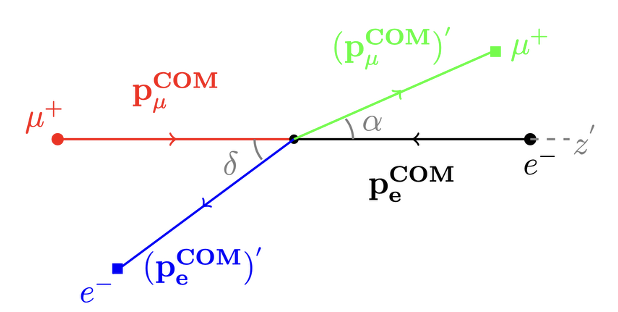}}
\caption{\justifying The $\mu e^- \to \mu e^- Z^\prime$ process depicted in (a) lab and (b) center-of-mass (COM) frames.}
\label{fig:frames}
\end{figure}

\begin{figure}[t]
\urulex{1pt}
\captionof{algorithm}{Efficient $\mu e^- \to \mu e^- Z^\prime$ event generation}
\label{alg:interp}
\drulex{1pt}
\begin{algorithmic}
\Function{GenerateMuEZprime}{$E_\mu$}
\State $E_{\mu1}, E_{\mu2}, \sigma_1, \sigma_2, H_1, H_2 \gets \Call{AdjacentGridPoints}{E_\mu}$;
\If{$E_\mu$ is out of the grid range}
\State \Return no $\mu e^- \to \mu e^-Z^\prime$ happens;
\EndIf
\State
$w_1 \gets \frac{E_{\mu} - E_{\mu2}}{E_{\mu1} - E_{\mu2}}$,
$w_2 \gets \frac{E_{\mu1} - E_{\mu}}{E_{\mu1} - E_{\mu2}}$;
\State $\sigma \gets w_1\sigma_1 + w_2\sigma_2$;
\If{$\Call{Random}{0, 1} < w_1$}
\State $\vec p_{\mu}, \vec p_e, \vec p_{miss} \gets H_1.\Call{Sample}{\null}$;\Comment{LAB}
\Else
\State $\vec p_{\mu}, \vec p_e, \vec p_{miss} \gets H_2.\Call{Sample}{\null}$;
\EndIf
\State $\vec p_{\mu}, \vec p_e, \vec p_{miss},  \gamma, \beta, \alpha, \phi \gets \Call{Kinematics}{E_\mu}$;\Comment{COM}
\State
$p_{\mu x} \gets p_{\mu}\sin\alpha\cos\phi$,
$p_{\mu y} \gets p_{\mu}\sin\alpha\sin\phi$;
\State
$p_{\mu z} \gets p_{\mu}\cos\alpha$;
\State
$p_{ex} \gets -p_{\mu x}-p_{miss\_x}$,
$p_{ey} \gets -p_{\mu y}-p_{miss\_y}$;
\State
$p_{ez} \gets -p_{\mu z}-p_{miss\_z}$;
\State
$p_{(\mu /e) x\_out} \gets p_{(\mu /e)x}$;\Comment{Boost}
\State
$p_{(\mu /e) y\_out} \gets p_{(\mu /e)y}$;
\State $p_{(\mu /e) z\_out} \gets \gamma\left(p_{(\mu /e)z} + \beta E_{(\mu /e)}/2\right)$;
\State $\vec p_{\mu \_out} \gets \Call{ThreeVector}{p_{\mu x}, p_{\mu y}, p_{\mu z}}$;
\State $\vec p_{\mu \_out} \gets \vec p_{\mu \_out}.\Call{RotateZAxisTo}{\hat p_\mu}$;
\State $\vec p_{e \_out} \gets \Call{ThreeVector}{p_{e x}, p_{e y}, p_{e z}}$;
\State $\vec p_{e \_out} \gets \vec p_{e \_out}.\Call{RotateZAxisTo}{\hat p_e}$;
\State \Return $\sigma, \vec p_{\mu \_out}, \vec p_{e\_out}$;
\EndFunction
\end{algorithmic}
\drulex{1pt}
\end{figure}


Interactions between the incoming muons with the material in the detector are simulated by GEANT4 11.2.2 \cite{Allison:2016lfl,Allison:2006ve,GEANT4:2002zbu}, where the $\mu e^- \to \mu e^-  Z^\prime$ process is implemented as a discrete process \cite{GEANT4:2023toolkit} following Algorithm~\ref{alg:interp}. Each event is classified as a signal (background) event if the $\mu e^- \to \mu e^- Z^\prime$ process does (does not) take place during the muon’s entire trajectory through the detector. For every combination of $m_{Z^\prime}$ and $E_\mu$, a global normalization factor is applied to the cross section so that the signal fraction remains at the level of $10^{-7}$–$10^{-3}$, ensuring high simulation efficiency while keeping multiple-scattering effects negligible. To maintain appropriate step granularity for the rare signal process, a dedicated routine is introduced to constrain the maximum step length to $10^{-3}$ times the scaled mean free path. This implementation achieves fine spatial resolution for the $\mu e^- \to \mu e^- Z^\prime$ interaction without compromising the overall computational performance. The signal process is integrated into the standard GEANT4 inclusive physics list \texttt{FTFP\_BERT}~\cite{GEANT4:2023physics}, which models the full set of hadronic and electromagnetic background interactions relevant to the detector environment.

The detector geometry in the simulation follows a modular design, as illustrated in Fig.~\ref{fig:pkmuonz}. The target composed of graphite has a transverse size of $100 \times 100~\mathrm{mm^2}$ and a thickness of $20~\mathrm{mm}$, providing a moderate electron density and a low muon--nuclear interaction probability. This allows efficient signal production while controlling background interactions.

Downstream of the target, the tracking system employs silicon strip detectors arranged in multiple configurations. Each module contains $1000$ strips of $0.1~\mathrm{mm} \times 100~\mathrm{mm} \times 320~\mu\mathrm{m}$ forming an X--Y readout pair. Two orthogonal planes are separated by a $1.8~\mathrm{mm}$ air gap to accommodate support structures and readout electronics. In addition to the standard X--Y planes, rotated modules (labeled as U and V) at $45^\circ$ and $135^\circ$ around the beam axis are installed to provide multiple projections, improving three-dimensional track reconstruction and resolving pattern ambiguities. This design is simplified from CMS 2S tracker module which is also used in MUonE experiment~\cite{Zoi:2024W1}. Modules are grouped along the beamline with alternating short and long gaps of $200~\mathrm{mm}$ and $500~\mathrm{mm}$, corresponding to intra-group and inter-group separations. This provides sufficient lever arms for accurate reconstruction of both incoming and outgoing leptons. 

\section{Event Reconstruction and selection}\label{sec:reco}

Track reconstruction proceeds in two stages. Upstream hits in the X--Y modules are converted to physical positions using the known strip pitch and layer orientation. A straight-line fit through the first four layers reconstructs the incoming muon trajectory, providing an initial estimate for the scattering vertex. Downstream reconstruction considers all possible X--Y hit pairs in the last six layers. Each candidate pair is validated against hits in the rotated U--V modules. Predicted positions in the U--V planes are transformed back using the known rotation angles ($45^\circ/135^\circ$) and compared with recorded hits. The combination with the best match defines the outgoing three-dimensional tracks.

Events are classified as high-quality events if each upstream layer contains exactly one hit and each downstream layer contains exactly two hits, ensuring unambiguous track patterns. The scattering vertex is determined by minimizing the distance between the outgoing tracks along the beam axis. Outgoing particle energies are recorded from MC at the first entry into the fifth silicon layer, immediately downstream of the target, capturing each particle's energy once. The reconstructed vertex is confirmed to coincide with the target position ($Z=-100~\mathrm{mm}$), and the resulting angular distributions show clear separation between small-angle muon tracks ($\leq 0.01~\mathrm{rad}$) and larger-angle electron tracks ($\leq 0.1~\mathrm{rad}$), allowing effective background suppression. The combination of precise hit selection, pattern matching, multiple projections, high spatial granularity, and fine step size ensures robust and efficient reconstruction of $\mu e^- \rightarrow \mu e^- Z'$ events.

\begin{figure}[t]
\subfloat[$E_\mu$ vs $\left<\vec{p}_0, \vec{p}_1\right>$]{\includegraphics[width=.5\columnwidth]{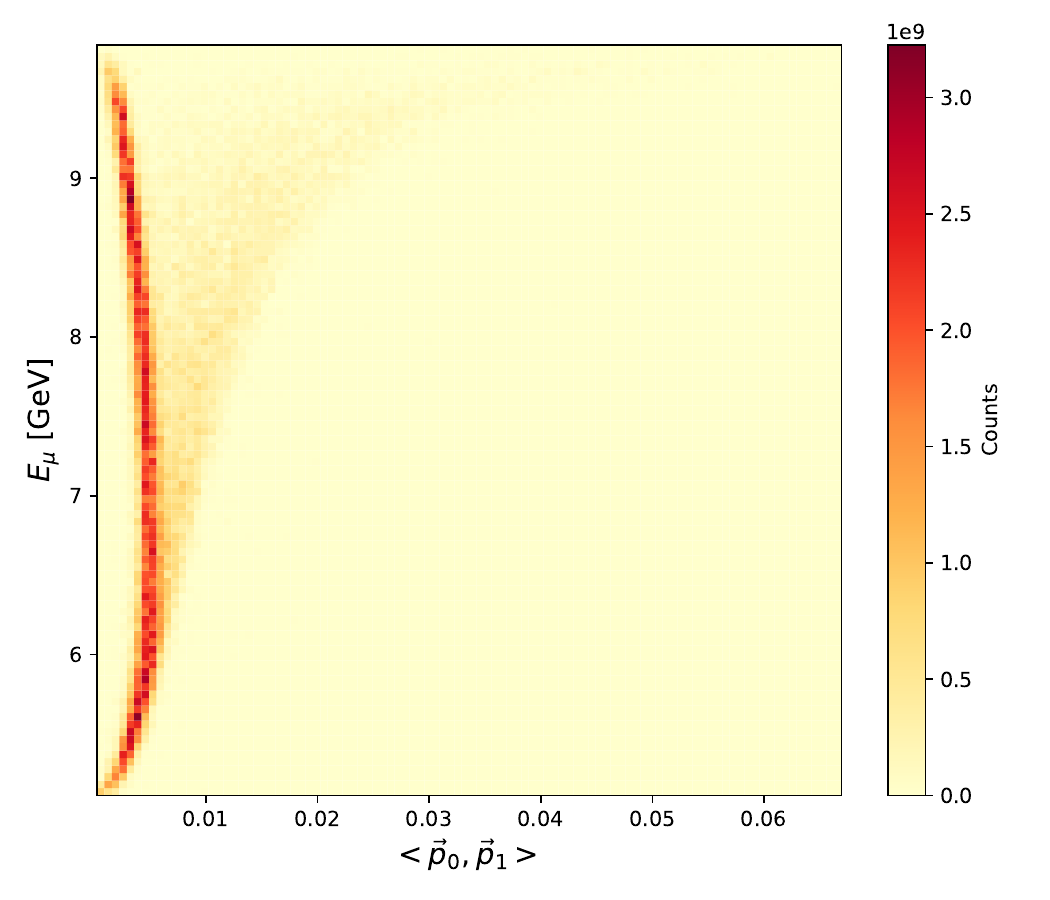}}
\subfloat[$E_\mu$ vs $\left<\vec{p}_0, \vec{p}_2\right>$]{\includegraphics[width=.5\columnwidth]{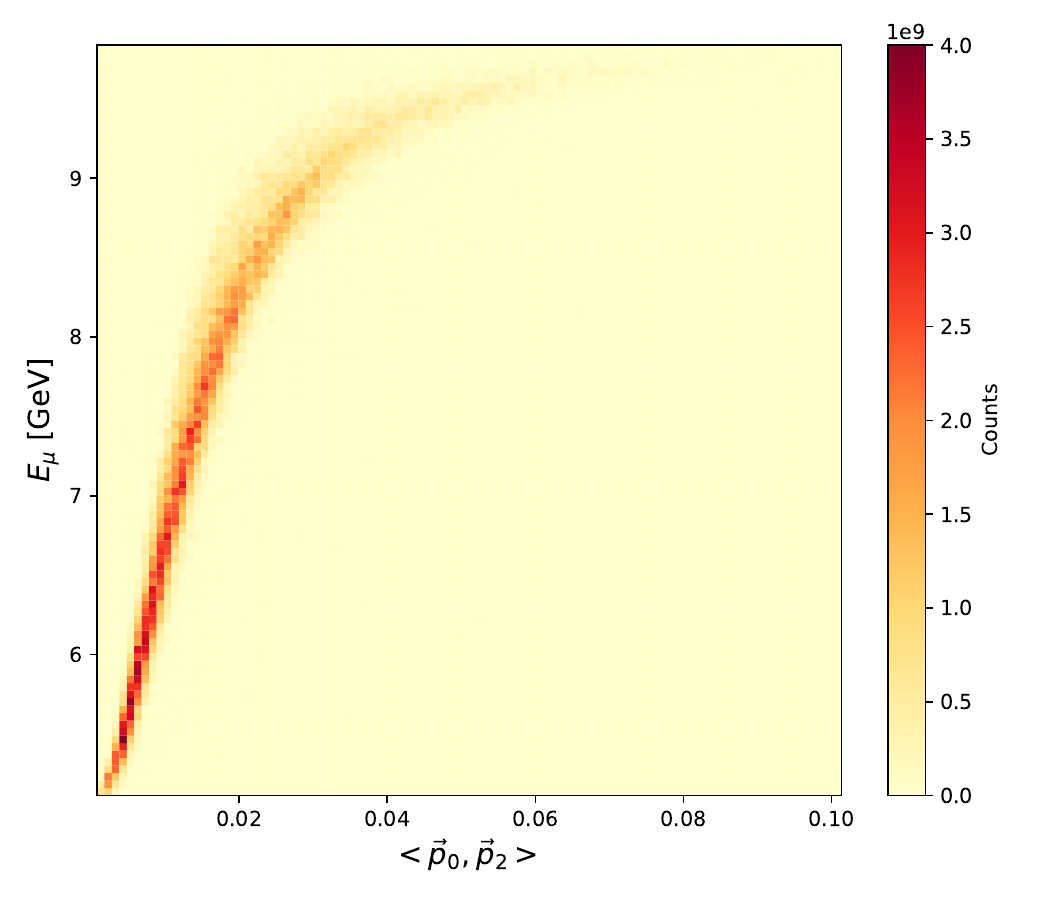}}

\subfloat[$E_e$ vs $\left<\vec{p}_0, \vec{p}_1\right>$]{\includegraphics[width=.5\columnwidth]{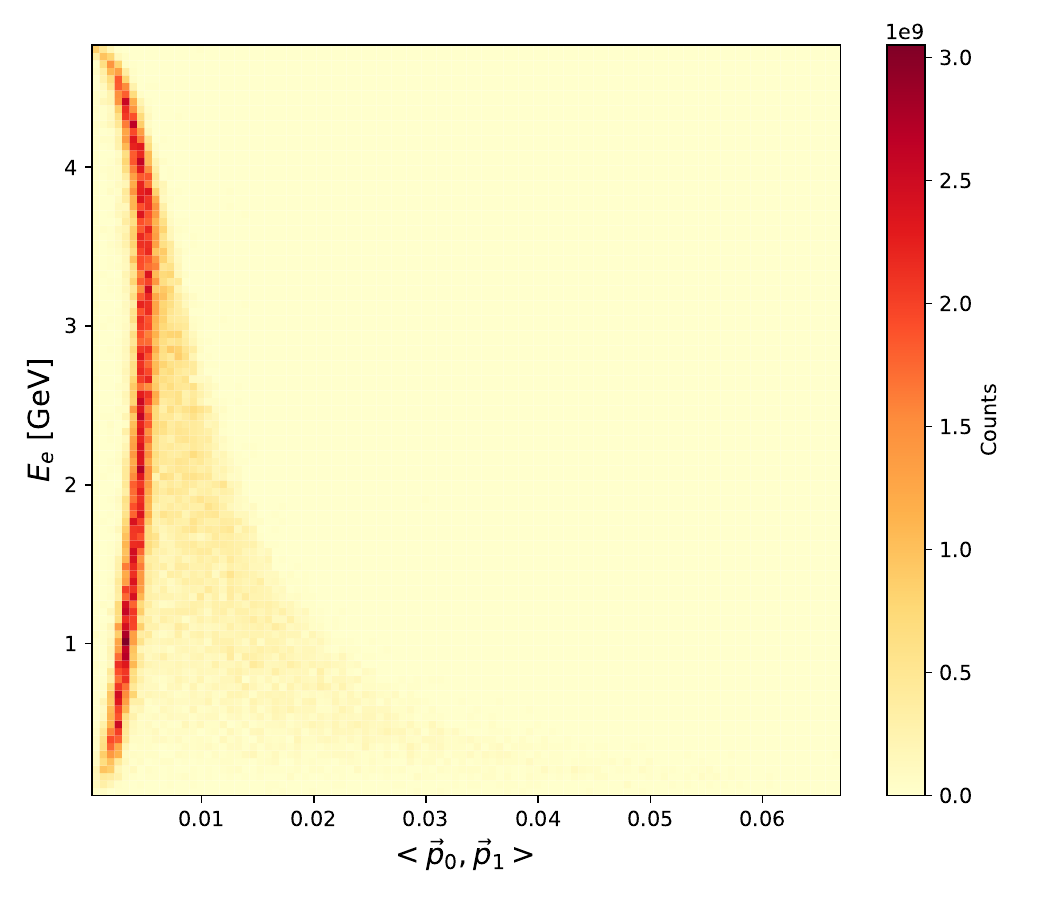}}
\subfloat[$E_e$ vs $\left<\vec{p}_0, \vec{p}_2\right>$]{\includegraphics[width=.5\columnwidth]{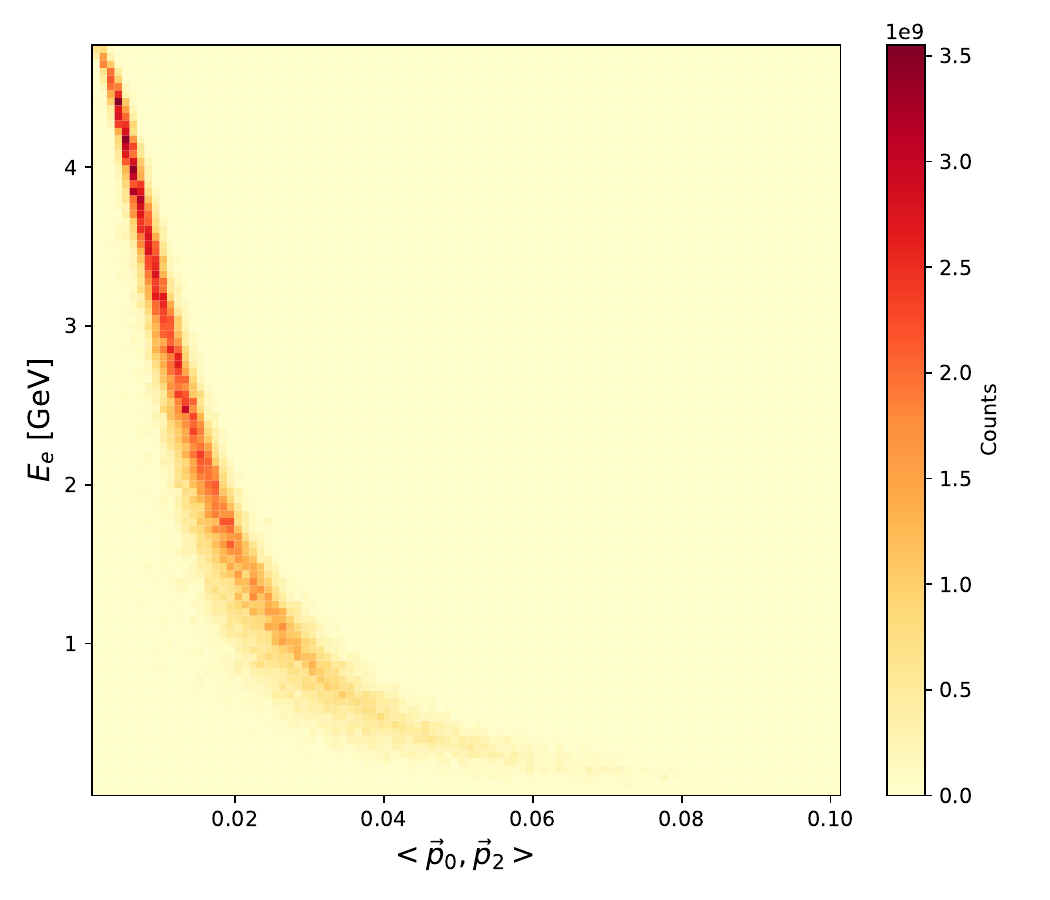}}

\caption{\justifying Correlation distributions for the signal process, illustrating the relations between particle energies and momentum vectors for the case of a 10~GeV muon beam and a $Z^\prime$ mass of 30~MeV. The target is a 20~mm graphite block. The yields are normalized to $2.5\times10^{14}$ muons on target. 
(a) Correlation between $E_\mu$ and $\left<\vec{p}_0, \vec{p}_1\right>$; 
(b) correlation between $E_\mu$ and $\left<\vec{p}_0, \vec{p}_2\right>$; 
(c) correlation between $E_e$ and $\left<\vec{p}_0, \vec{p}_1\right>$; 
(d) correlation between $E_e$ and $\left<\vec{p}_0, \vec{p}_2\right>$.
}
\label{fig:angle_sig}
\end{figure}

\begin{figure}[t]
\subfloat[$E_\mu$ vs $\left<\vec{p}_0, \vec{p}_1\right>$]{\includegraphics[width=.5\columnwidth]{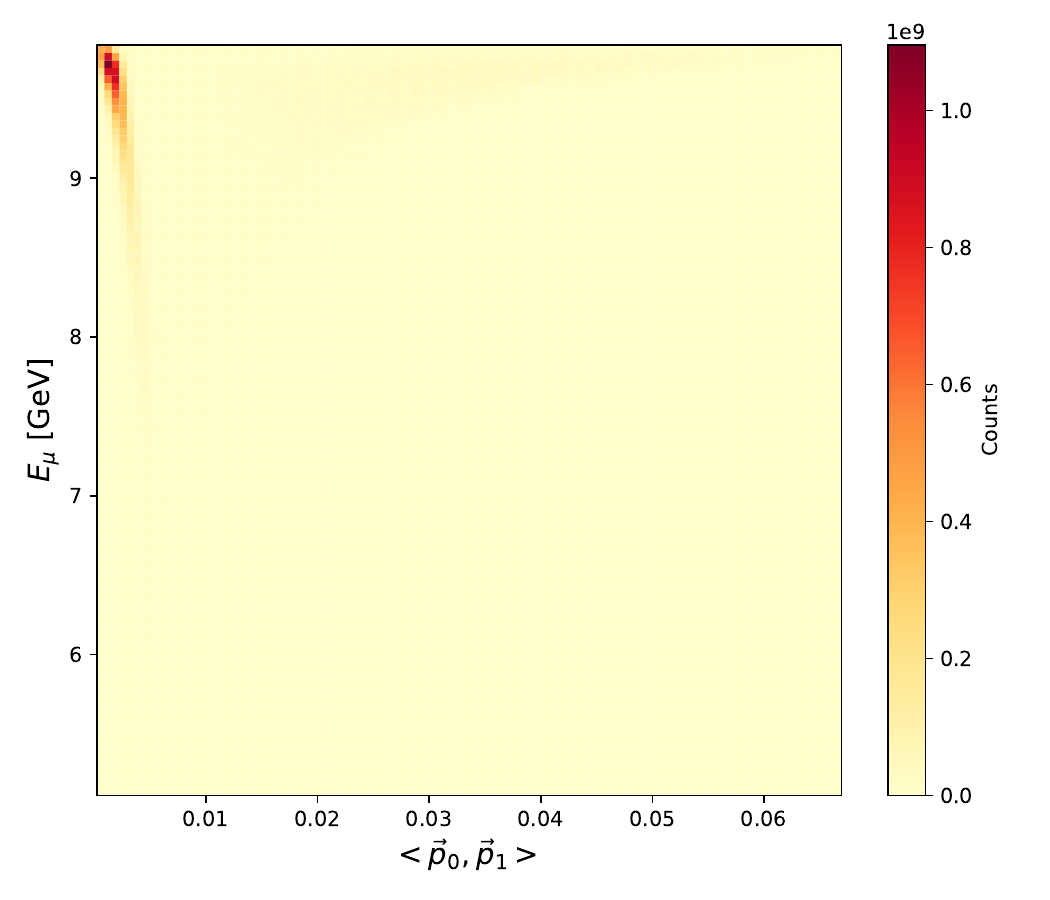}}
\subfloat[$E_\mu$ vs $\left<\vec{p}_0, \vec{p}_2\right>$]{\includegraphics[width=.5\columnwidth]{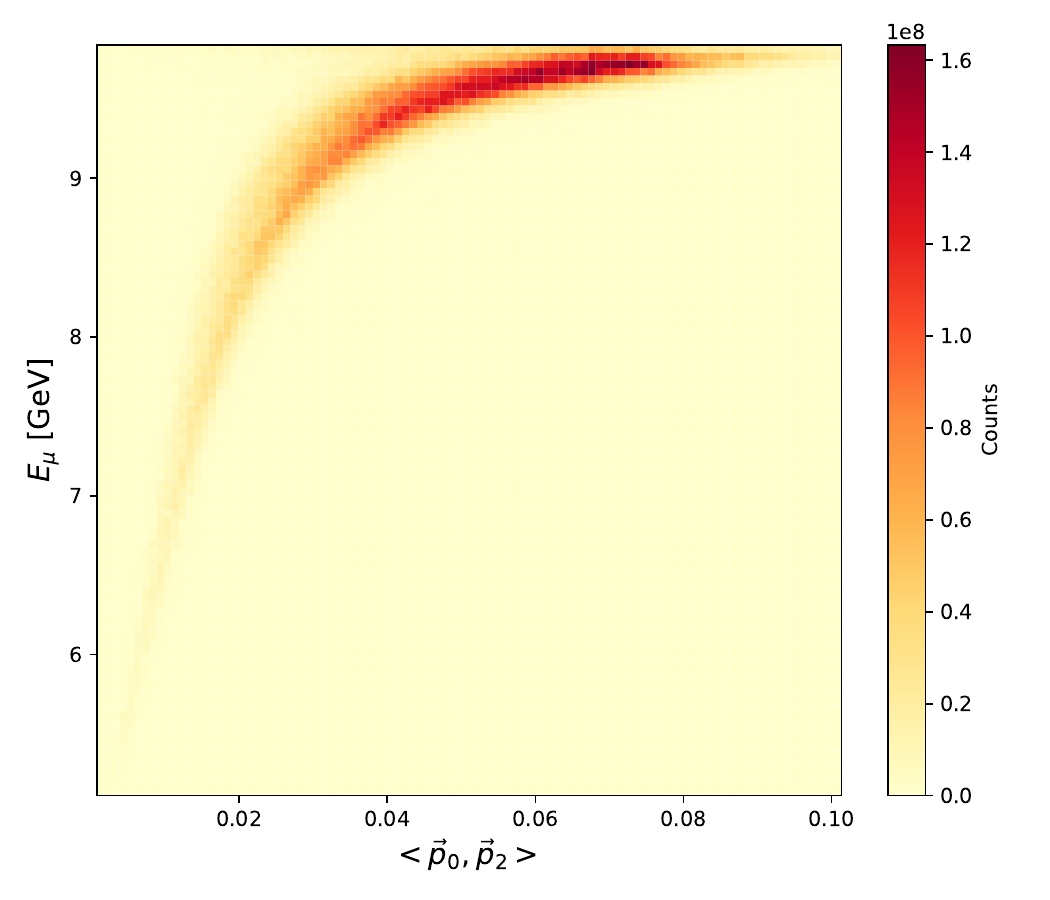}}

\subfloat[$E_e$ vs $\left<\vec{p}_0, \vec{p}_1\right>$]{\includegraphics[width=.5\columnwidth]{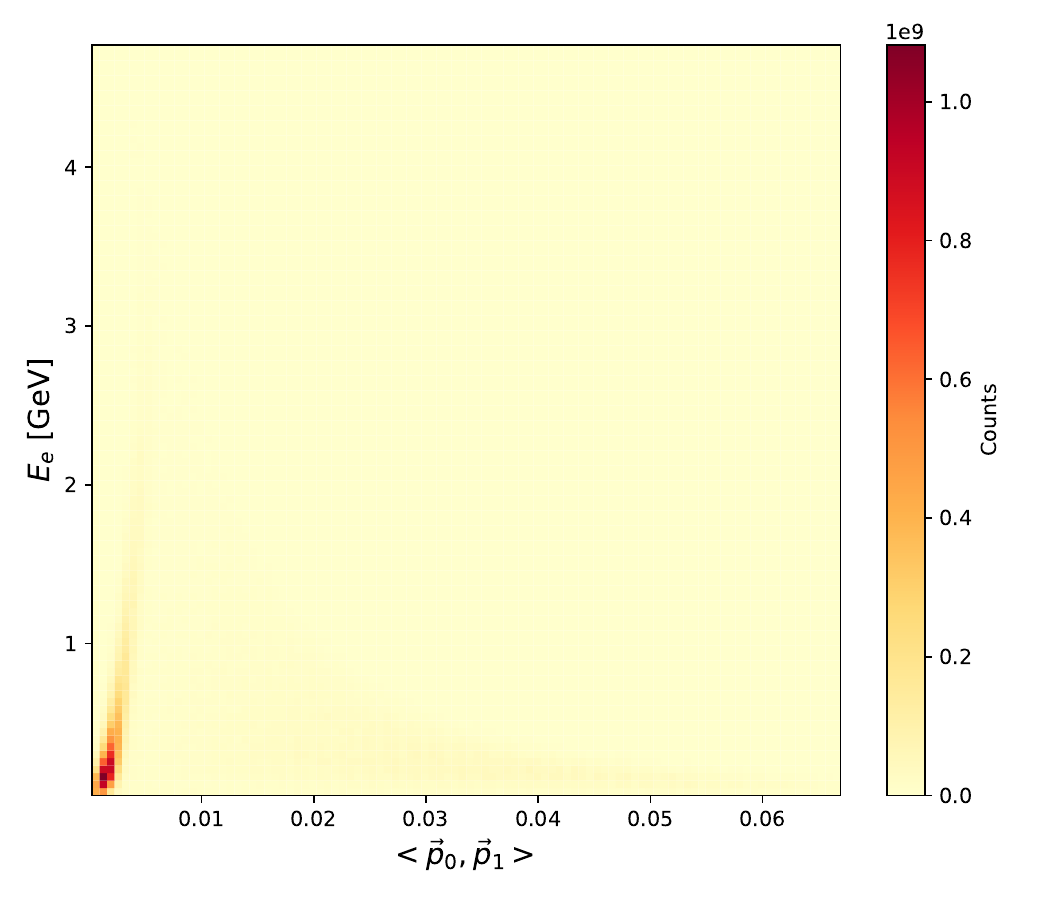}}
\subfloat[$E_e$ vs $\left<\vec{p}_0, \vec{p}_2\right>$]{\includegraphics[width=.5\columnwidth]{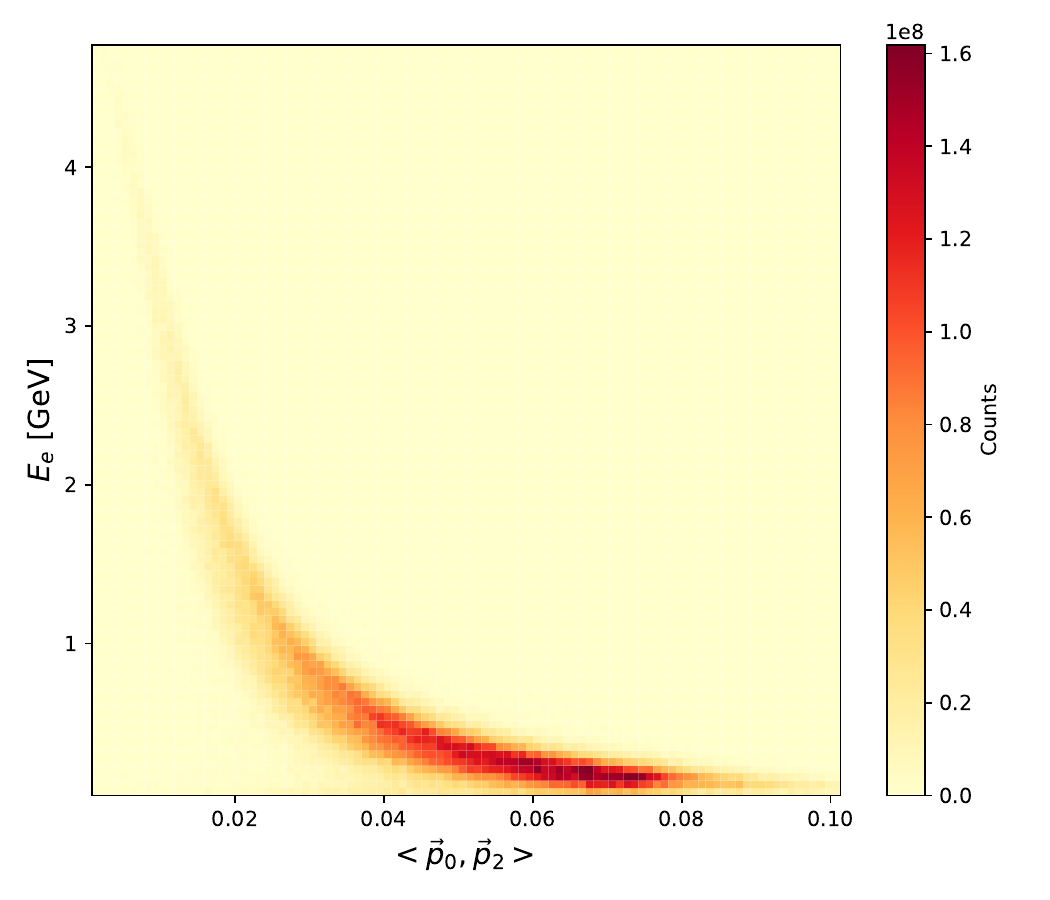}}

\caption{\justifying Correlation distributions for the inclusive background, illustrating the relations between particle energies and momentum vectors for the case of a 10~GeV muon beam. The target is a 20~mm graphite block. The yields are normalized to $2.5\times10^{14}$ muons on target.
These four distributions correspond directly to the four signal distributions shown in Fig.~\ref{fig:angle_sig}.}
\label{fig:angle_bkg}
\end{figure}

FIG.~\ref{fig:angle_sig} shows the two-dimensional correlations for the signal process, illustrating how the muon energy $E_\mu$ and the electron energy $E_e$ relate to the momentum vector correlations $\langle \vec{p}_0, \vec{p}_1 \rangle$ and $\langle \vec{p}_0, \vec{p}_2 \rangle$, where the momenta of the three tracks are denoted by $\vec{p}_0$, $\vec{p}_1$ and $\vec{p}_2$, corresponding to incoming muons, outgoing muons and outgoing electrons. The results correspond to a 10~GeV muon beam and a $Z^\prime$ mass of 30~MeV, with a 20~mm graphite target. $10^{9}$ events are generated for each plot and scaled to $2.5\times10^{14}$ muons on target, corresponding to a one-year run estimated from a beam rate of $8.2\times10^{6}~\mu$/s over $3.1\times10^{7}$~s, with a scale factor of $10^{4}$ applied. FIG.~\ref{fig:angle_bkg} (a)--(d) present the background distributions for the case of a 10 GeV muon beam. Compared with the signal, the background shows that $E_\mu$ is more concentrated in the high-energy region (mostly above 9~GeV), while $E_e$ is concentrated in the low-energy region (mostly below 1~GeV), reflecting the strong correlation between the two due to momentum conservation. In addition, the scattering angles of both the muon and the electron in the background process tend to cluster at larger angles.

To optimize signal selection for the $\mu e^- \to \mu e^- Z^\prime$ process, we employ a Boosted Decision Tree (BDT) using the \texttt{XGBoost}~\cite{Chen_2016} framework. The BDT was trained on a set of kinematic variables that are sensitive to the presence of the signal process, including the energy of electron $E_e$,  scattering angles ($\theta_\mu$, $\theta_e$), the azimuthal angle difference $\Delta\theta$. The training dataset consisted of labeled signal and background events, with 80\% used for training and 20\% for validation to prevent overfitting. The BDT hyperparameters shown in Table~\ref{tab:bdt_params} were tuned to achieve an optimal balance between signal-background discrimination and generalization. The tuning was guided by monitoring the ROC curve and the corresponding AUC, ensuring a high true positive rate while keeping the false positive rate low. Overtraining was further controlled by comparing BDT score distributions between training and validation samples, guaranteeing reliable performance on unseen events. This ensured maximal signal-background separation without compromising performance on unseen events. 

\begin{table}[h!]
\centering
\caption{BDT hyperparameters used in the analysis.}
\label{tab:bdt_params}
\begin{tabular}{lc}
\hline
\hline
\textbf{Hyperparameter} & \textbf{Value} \\
\hline
Maximum tree depth & 4 \\
Learning rate & 0.1 \\
Maximum number of boosting rounds & 1000 \\
LAMBDA & 10 \\
\hline
\hline
\end{tabular}
\end{table}

After training, 
the BDT output score was used to define the final event selection. For each mass point, a separate BDT was trained using the corresponding events and an optimal threshold on the BDT score was determined to maximize the signal significance, effectively suppressing the dominant background contributions while retaining a high fraction of the signal. To further enhance the background rejection and improve the overall selection purity, an additional final cut was applied on top of the BDT selection to further refine the event sample:
\begin{flalign*}
&\text{(i)} \quad E_e \ge 4.25~\text{GeV},&\\
&\text{(ii)} \quad \theta_e \le 0.02~\text{rad}, &\\
&\text{(iii)} \quad \theta_\mu \ge 0.004~\text{rad}. &
\end{flalign*}

This final cut ensures that the remaining events are enriched in the $\mu e^- \to \mu e^- Z^\prime$ signal.

\section{Results}\label{results}

\begin{figure}[H]
\includegraphics[width=1.0\columnwidth]{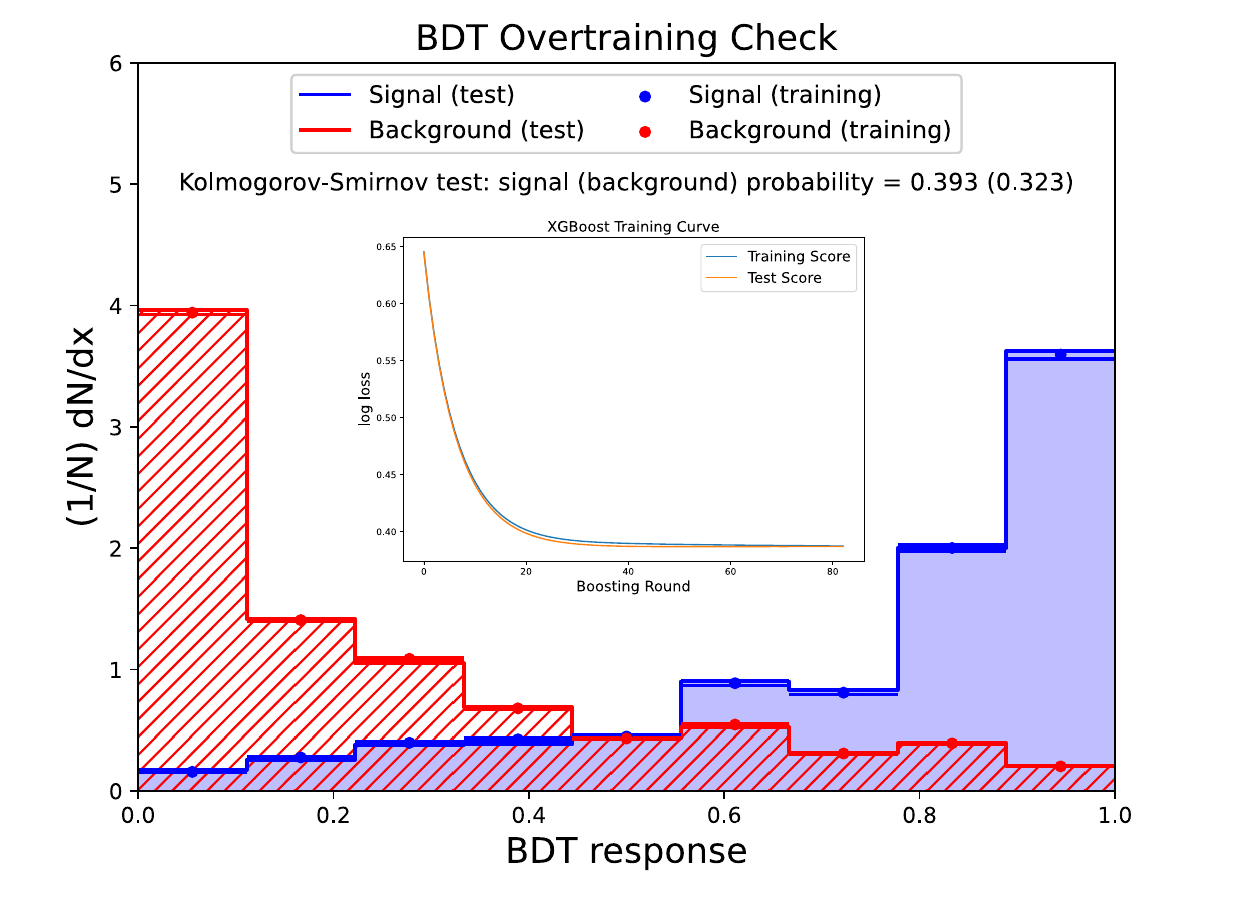}
\caption{\justifying BDT overtraining check comparing the output score distributions of signal (blue) and background (red) for both training and test samples with $E_\mu = 10~\mathrm{GeV}$ and $m_{Z^{\prime}} = 30~\mathrm{MeV}$. And evolution of the BDT training and test log loss as a function of boosting rounds for the $\mu e^- \to \mu e^- Z^\prime$ process.}
\label{fig:overtraining}
\end{figure}

Figure~\ref{fig:overtraining} shows the performance of the BDT training for the $\mu e^- \to \mu e^- Z^\prime$ process at $E_\mu = 10~\mathrm{GeV}$ and $m_{Z^{\prime}} = 30~\mathrm{MeV}$.
The evolution of the training and test losses as a function of the boosting rounds indicates a rapid initial decrease followed by smooth convergence for both datasets, suggesting stable training behavior without signs of overfitting. The close agreement between the training and test loss curves further confirms that the BDT model generalizes well to unseen data.

For the overtraining check, the BDT output score distributions of the signal and background samples are compared between the training and test datasets. The blue (signal) and red (background) histograms are clearly separated, indicating that the classifier effectively distinguishes between the two classes. The Kolmogorov–Smirnov (K–S) test probabilities for the signal (0.393) and background (0.323) samples confirm the absence of significant overtraining, validating the robustness of the trained model. Overall, the BDT demonstrates excellent discriminating power between the $\mu e^- \to \mu e^- Z^\prime$ signal and the corresponding background processes.

In total, 100 mass points were analyzed for each beam energy. After training, the optimal BDT score threshold was automatically determined to maximize the signal significance, followed by the application of the final event selection described in the previous section.

After the selection, as suggested in Refs. \cite{Li:2023lin,Cowan:2010js}, the test statistics $Z$ for 95\% confidence level upper limit is utilized to describe the expected sensitivity of the proposed experiment to the $\mu e^- \to \mu e^-Z^{\prime}$ process:
\begin{equation}
Z = 2\left(\mu s + b\log\left(b / (\mu s + b)\right)\right) \sim \chi^2(1),
\end{equation}
where $s$ and $b$ are, respectively, the post-selection signal and background yields normalized to one year, and $\mu$ is the signal strength to be limited by the upper 5\% point of the $\chi^2(1)$ distribution. The result is finally translated into the limit on $g_{Z^{\prime}}$ by multiplying it by the global cross section scale factor applied to the corresponding simulation run with $g_{Z^{\prime}} = 1$ as initial setting. 

\begin{figure}[H]
\includegraphics[width=1.0\columnwidth]{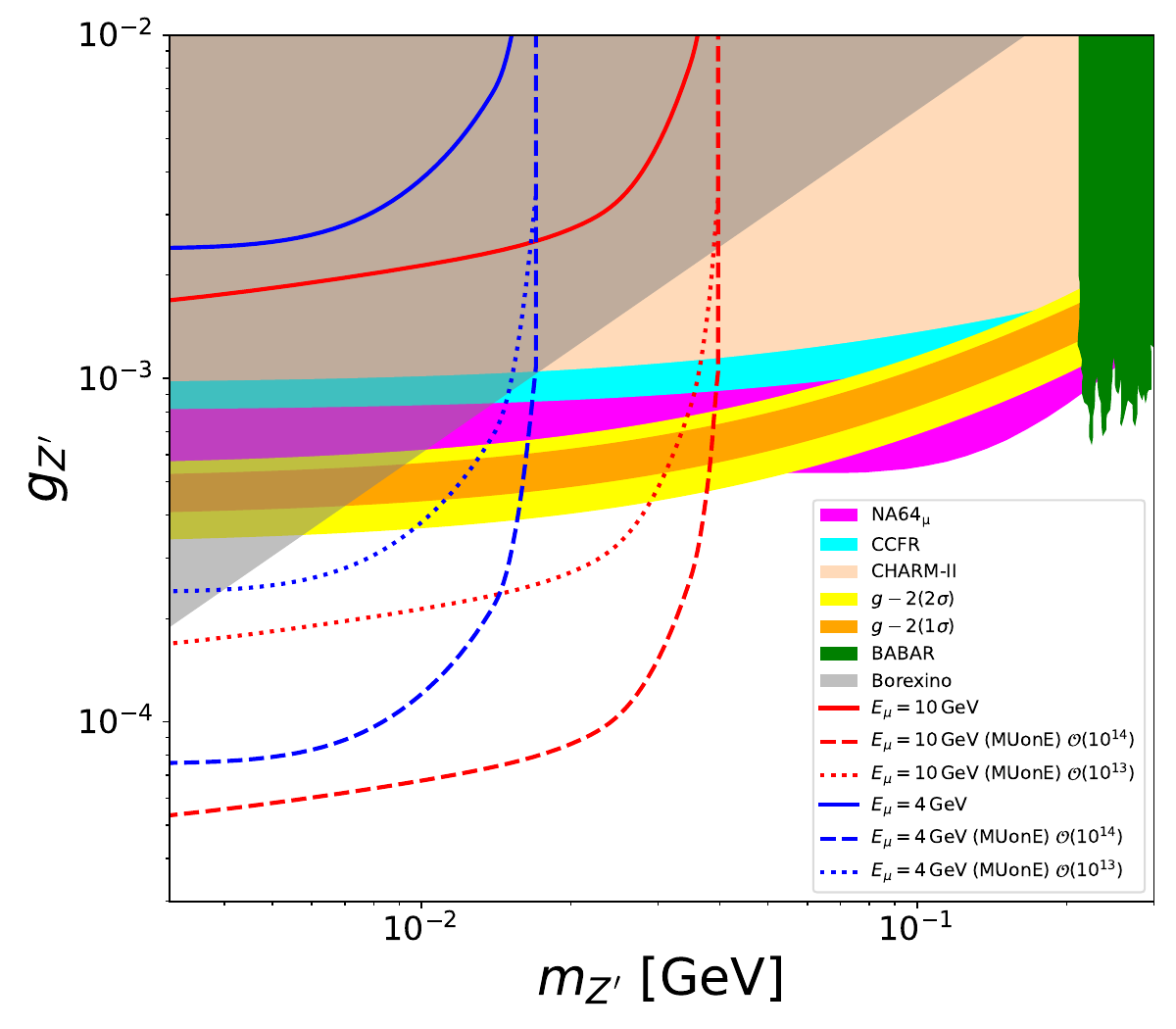}
\caption{\justifying 
Upper limits on the coupling $g_{Z^\prime}$ as a function of $Z^\prime$ mass $m_{Z^\prime}$. The solid red and blue lines represent the limits obtained for a muon beam energy of 10~GeV and 4~GeV, respectively. The dashed red and blue lines indicate the expected upper limits in the MUonE experiment~\cite{Asai:2021wzx} , assuming 40 stations and 10~cm thick lead targets. The red and blue dotted lines represent the MUonE expected limits under a conservative estimate of $10^{13}$ MOT for a one-year run. The shaded areas in light brown, light blue, purple, green and gray are excluded by CHARM-II~\cite{GEIREGAT1990271}, CCFR~\cite{PhysRevLett.66.3117}, NA64$_\mu$~\cite{NA64:2024nwj},  BABAR~\cite{PhysRevD.94.011102} and Borexino~\cite{PhysRevLett.107.141302}, respectively. The orange and yellow bands show the regions favored by the muon $g-2$ anomaly at the 1$\sigma$ and 2$\sigma$ levels, respectively.
}
\label{fig:UL}
\end{figure}

FIG.~\ref{fig:UL} presents the expected $95\%$ confidence level (C.L.) upper limits on the coupling strength $g_{Z^{\prime}}$ as a function of the $Z'$ boson mass $m_{Z^{\prime}}$. The solid red and blue curves correspond to our simulated results for muon beam energies of 10~GeV and 4~GeV, respectively.
The curves were smoothed using a cubic smoothing spline method to account for detector response effects and reconstruction efficiency fluctuations, ensuring a physically continuous representation of the limits. The shaded regions represent the current exclusion limits from existing experiments: CHARM-II~\cite{GEIREGAT1990271} (light brown), CCFR~\cite{PhysRevLett.66.3117} (light blue), NA64$_\mu$~\cite{NA64:2024nwj} (purple), BABAR~\cite{PhysRevD.94.011102} (green) and Borexino~\cite{PhysRevLett.107.141302} (gray). The orange and yellow bands indicate the regions favored by the muon $g-2$ anomaly at the $1\sigma$ and $2\sigma$ confidence levels, respectively.

The dashed red and blue curves represent the projected sensitivities that could be achieved if the same search were performed using the multi-station telescope setup, such as MUonE experiment with 40 stations~\cite{Asai:2021wzx}. In this scenario, each station would be equipped with one target, and the original graphite target could be replaced with a 10~cm thick lead target. This configuration would increase the expected signal rate by roughly three orders of magnitude due to the larger target thickness and higher electron density, assuming that each station can achieve the same level of background suppression efficiency as demonstrated in our analysis. Under these conditions, the MUonE experiment could achieve unprecedented sensitivity to the $Z^{\prime}$ coupling, particularly in the low-mass region around $m_{Z^{\prime}} \sim 10$~MeV, where the expected upper limits surpass all current experimental constraints. Considering practical limitations such as the purity of the muon beam, we adopt a conservative estimate in which the muon on target (MOT) for a one-year run is reduced by one order of magnitude, i.e., $10^{13}$. The corresponding MUonE expected limits are shown in the figure as red and blue dotted lines. Even under this conservative assumption, the projected sensitivity remains superior in the mass region of interest around $m_{Z^{\prime}} \sim 10$~MeV, demonstrating that a dedicated MUonE-based search could provide some of the world’s most stringent bounds on the $Z^{\prime}$ dark boson. 

\section{Summary and Outlook}\label{sec:summary}

In this work, we have proposed a muon on-target experiment based on the HIAF muon beam to search for light dark sector particles within the $L_\mu - L_\tau$ model, focusing on the process $\mu e^- \to \mu e^- Z^\prime$ with an invisibly decaying $Z^\prime$. Using the high-intensity, tunable 1--10~GeV muon beam available at HIAF, our study demonstrates that this energy regime provides excellent sensitivity to light $Z^\prime$ bosons, particularly in the $\mathcal{O}(10~\mathrm{MeV})$ mass range.

A full simulation framework was developed based on precomputed kinematic grids using MadGraph5\_aMC@NLO and GEANT4, enabling efficient event generation across multiple beam energies and $Z^\prime$ mass hypotheses. The detector design follows the PKMu concept, featuring silicon tracking modules before and after the target, allowing accurate vertex reconstruction and precise angular measurements. Through a dedicated reconstruction algorithm and BDT-based selection, we achieved a strong separation between signal and background, leading to competitive upper limits on the coupling strength $g_{Z^\prime}$.


Our projected 95\% confidence level limits show that, for a 10~GeV muon beam and a 20~mm graphite target, the experiment can reach sensitivity to $g_{Z^\prime}$ at the $10^{-3}$ level. Furthermore, if implemented in the MUonE setup with 40 stations and 10~cm lead targets, the expected signal yield could increase by nearly three orders of magnitude, achieving world-leading sensitivity in the $m_{Z^\prime} \sim 10~\mathrm{MeV}$ region. In conclusion, the proposed HIRIBL-PKMu experiment provides a promising new approach for probing light dark forces. Our work thus lays an essential foundation for future precision muon experiments at HIAF and opens a new avenue toward exploring the dark sectors of particle physics.

\begin{acknowledgments}
This work is supported in part by the National Natural Science Foundation of China under Grants No. 12325504, and No. 12061141002. The work of C. Z. was supported by the Leverhulme Trust, LIP-2021-014.
\end{acknowledgments}

\appendix

\section{Kinematic Solution and Reference Frame Transformation in the $\mu e^- \rightarrow \mu e^- Z^{\prime}$ Process}
\label{sec:kinematic}

For the $\mu e^- \rightarrow \mu e^- Z^{\prime}$ process, the kinematic reconstruction implemented in the simulation follows the procedure summarized below. 
In the laboratory (lab) frame, an incoming muon with total energy $E_\mu$ and mass $m_\mu$ scatters off a stationary electron of mass $m_e$, producing an outgoing muon, an outgoing electron, and a missing boson $Z'$ with four-momentum $p_{Z^{\prime}} = (E_{Z^{\prime}}, \vec{p}_{Z^{\prime}})$. 

The center-of-mass (COM) energy of the system is given by
\begin{equation}
E_{\mathrm{COM}} = \sqrt{m_\mu^2 + m_e^2 + 2E_\mu m_e} ,
\end{equation}
corresponding to the Lorentz boost parameters from the COM frame to the lab frame:
\begin{equation}
\gamma = \frac{E_\mu + m_e}{E_{\mathrm{COM}}} \qquad , 
\beta = \frac{\sqrt{(E_\mu + m_e)^2 - E_{\mathrm{COM}}^2}}{E_\mu + m_e}.
\end{equation}

These define the longitudinal Lorentz transformation as
\begin{equation}
p_z = \gamma (p'_z + \beta E'), \qquad 
E = \gamma (E' + \beta p'_z).
\end{equation}
where primed quantities denote momenta in the COM frame.

In the COM frame, the outgoing muon momentum direction is parameterized by a polar angle $\alpha$ and an azimuthal angle $\phi$, with the unit vector
\begin{equation}
\hat{n}_\mu' = 
(\sin\alpha \cos\phi,\, \sin\alpha \sin\phi,\, \cos\alpha).
\end{equation}

Momentum conservation in the three-body final state introduces a correlation between the outgoing muon and the missing $Z^{\prime}$ momentum. 
Denoting the magnitude of the missing momentum as $p_{Z^{\prime}} = |\vec{p}_{Z^{\prime}}|$, and its direction as $\hat{p}_{Z^{\prime}} = \vec{p}_{Z^{\prime}}/p_{Z^{\prime}}$, one can construct the scalar products
\begin{align}
A &= p_{Z^{\prime}} \, \hat{p}_{Z^{\prime}} \cdot \hat{n}_\mu'\nonumber ,\\
B &= \frac{E_{\mathrm{COM}}^2 + m_\mu^2 + m_{Z^{\prime}}^2 - m_e^2}{2}
     - E_{\mathrm{COM}} E_{Z^{\prime}}\nonumber ,\\
C &= E_{Z^{\prime}} - E_{\mathrm{COM}} .
\end{align}

These quantities lead to a quadratic constraint on the magnitude $p_\mu'$ of the outgoing muon momentum in the COM frame:
\begin{multline}
(A^2 - C^2)\, p_\mu'^2 
- 2 A B\, p_\mu' 
+ (B^2 - C^2 m_\mu^2) = 0\\[-3pt]
\text{with } 
\Delta = (A B)^2 - (A^2 - C^2)(B^2 - C^2 m_\mu^2).
\label{disciminant}
\end{multline}

Two possible real solutions exist:
\begin{align}
p_{\mu,\pm}' = 
\frac{A B \pm \sqrt{\Delta}}
     {A^2 - C^2}.
\end{align}

Among them, the physically meaningful solution corresponds to the positive momentum value that satisfies energy conservation and kinematic consistency.

After the COM momenta are obtained, the outgoing muon and electron four-momenta are boosted back to the lab frame using the Lorentz transformation along the beam axis. 
The final-state three-momenta in the lab frame are then given by
\begin{align}
\vec{p}_\mu &= 
\begin{pmatrix}
p'_\mu \sin\alpha \cos\phi \\[1mm]
p'_\mu \sin\alpha \sin\phi \\[1mm]
\gamma (p'_\mu \cos\alpha + \beta E'_\mu)
\end{pmatrix}, \\
\vec{p}_e &= 
\begin{pmatrix}
-\,p'_\mu \sin\alpha \cos\phi - p_{Z^\prime}^x \\[1mm]
-\,p'_\mu \sin\alpha \sin\phi - p_{Z^\prime}^y \\[1mm]
-\,p'_\mu \cos\alpha - p_{Z^\prime}^z
\end{pmatrix}.
\end{align}

This procedure ensures full four-momentum conservation at every step and allows the event generator to reconstruct the three-body final state based on sampled angular variables $(\alpha, \phi)$ and the missing boson momentum $p_{Z^{\prime}}$.The discriminant form of Eq.~\ref{disciminant} guarantees that all generated events satisfy the physical on-shell conditions, while the Lorentz transformation preserves the spatial and temporal consistency of each simulated scattering. 
This algorithmic approach thus provides a numerically stable and physically accurate realization of the $\mu e^- \rightarrow \mu e^- Z^{\prime}$ process within the \textsc{Geant4} simulation framework.


\bibliographystyle{unsrt} 
\bibliography{apssamp}

\end{document}